\documentclass[aps,prl,showpacs,twocolumn,superscriptaddress,floatfix]{revtex4}

\usepackage{graphicx}% Include figure files
\usepackage{dcolumn}% Align table columns on decimal point
\usepackage{bm}% bold math
\usepackage{amsmath}
\usepackage{amssymb}
\usepackage{color}% color

\begin{document}

\renewcommand{\vec}[1]{\mbox{\boldmath $#1$}}

%\preprint{}

\title{Spin-orbit fluctuations in frustrated heavy-fermion metal LiV$_2$O$_4$}


\author{K. Tomiyasu}
\email[Electronic address: ]{tomiyasu@m.tohoku.ac.jp}
\affiliation{Department of Physics, Tohoku University, Aoba, Sendai 980-8578, Japan}
\author{K. Iwasa}
\affiliation{Department of Physics, Tohoku University, Aoba, Sendai 980-8578, Japan}
\author{H. Ueda}
\affiliation{Department of Chemistry, Kyoto University, Kyoto 606-8502, Japan}
\author{S. Niitaka}
\affiliation{RIKEN, 2-1 Hirosawa, Wako, Saitama 351-0198, Japan}
\author{H. Takagi}
\affiliation{Department of Physics, University of Tokyo, Hongo 113-0033, Japan}
\author{S. Ohira-Kawamura}
\affiliation{J-PARC Center, Japan Atomic Energy Agency, Tokai, Ibaraki 319-1106, Japan}
\author{T. Kikuchi}
\affiliation{J-PARC Center, Japan Atomic Energy Agency, Tokai, Ibaraki 319-1106, Japan}
\author{K. Nakajima}
\affiliation{J-PARC Center, Japan Atomic Energy Agency, Tokai, Ibaraki 319-1106, Japan}
\author{K. Yamada}
\affiliation{Institute of Materials Structure Science, High Energy Accelerator Research Organization, Oho, Tsukuba 305-0801, Japan}
%\thanks{}


\date{\today}

\begin{abstract}
Spin fluctuations were studied over a wide momentum ($\hbar Q$) and energy ($E$) space in the frustrated $d$-electron heavy-fermion metal LiV$_2$O$_4$ by time-of-flight inelastic neutron scattering. 
We observed the overall $Q$--$E$ evolutions near the characteristic $Q=0.6$ {\AA}$^{-1}$ peak and found another weak broad magnetic peak around 2.4 {\AA}$^{-1}$.  
The data are described by a simple response function, a highly itinerant magnetic form factor, and antiferromagnetic short-range spatial correlations, indicating that heavy-fermion formation is attributable to spin-orbit fluctuations with orbital hybridization. 
\end{abstract}

% up to 4
\pacs{71.27.+a, 75.25.-j, 75.40.Gb, 78.70.Nx}
%\keywords{geometrical frustration, inelastic neutron scattering, spin molecule, dimer}

\maketitle

%
%\section{Introduction}
%

% Quasiparticles and many-body problem
The many-body problem is central to modern condensed-matter physics, i.e., how does one describe a large number of intricately interacting particles in solids and liquids? The concept of quasiparticles constitutes the basis of this problem; a system can be successfully treated as a collection of independent quasiparticles~\cite{Landau_text}. Examples include heavy fermions (HF) in metals and Cooper pairs in superconductors, in which conduction electrons are coupled with spins and lattices. 

%LiV2O4, frustration
In 1997, the heaviest fermion system among $d$-electron systems, the metallic spinel LiV$_2$O$_4$ (nominally V$^{3.5+}$, 3$d^{1.5}$), was discovered~\cite{Kondo_1997}. The ratio of the heat capacity to temperature $C/T$ steeply increases with a large Sommerfeld coefficient $\gamma\simeq400$ mJ$\cdot$mol$^{-1}$$\cdot$K$^{-2}$ below the characteristic temperature $T^{*}\simeq20$ K~\cite{Urano_2000, Matsushita_2005, Das_2007}. This followed the report of another $d$-electron HF system, Y(Sc)Mn$_2$, with $\gamma\simeq140$ mJ$\cdot$mol$^{-1}$$\cdot$K$^{-2}$~\cite{Wada_1987}. In both $d$-electron systems, the magnetic atoms form a geometrically frustrated pyrochlore lattice, suggesting a close connection between the HF and frustration. 

%LiV2O4, spin
LiV$_2$O$_4$ exhibits a weak cusp in magnetic susceptibility at $T^{*}$ but no magnetic order at any measured temperature, indicating strong frustration~\cite{Urano_2000}. Instead, powder inelastic neutron scattering (INS), nuclear magnetic resonance (NMR), and muon spin resonance ($\mu$SR) detect spin fluctuations below $\sim80$ K down to 20 mK, which increase to antiferromagnetic (AF) short-range fluctuations described by $Q \simeq 0.6$ {\AA}$^{-1}$ below $T^{*}$~\cite{Lee_2001, Shimizu_2012, Kadono_2012}, where the magnitude of the momentum $p = \hbar Q$. The 0.6-{\AA}$^{-1}$ nesting structure is also obtained by band calculations~\cite{Yushankhai_2007}. 
%Gamma-gamma relation in SCR theory$B$+$i$b(Bspin dominating HF$B$,<(:6$5$l$?!#(B

%LiV2O4, charge
In addition, the electrical resistivity is metallic over the entire temperature range below room temperature and further decreases below $T^{*}$~\cite{Urano_2000}. This decrease is different from the Kondo upturn, which is the fingerprint of conventional $f$-electron HF systems based on the Kondo coupling between the localized $f$-electron momenta and the conduction electrons. Further, the optical conductivity suggests that LiV$_2$O$_4$ changes from a poor metal to a coherent Fermi-liquid metal around $T^{*}$ as the temperature decreases, as in the vicinity of a Mott insulator~\cite{Jonsson_2007}. Photoemission also resulted in a resonance peak in the electronic density of states at $\sim4$ meV above the Fermi level~\cite{Shimoyamada_2006}, which is also theoretically understood as the vicinity of the Mott insulator~\cite{Arita_2007}. 

% Motivation: requirements of detailed INS
Thus, HF formation likely originates not from the conventional Kondo effect but from another novel electron correlation effect. The 0.6-{\AA}$^{-1}$ AF spin fluctuations driven by frustration will play a key role in HF formation. However, the overall correlations of the spin fluctuations in a wide $(Q,E)$ space are still unclear, where $E$ denotes the energy. For example, many $Q$-dependent characteristic frequencies were reported around 0.6 {\AA}$^{-1}$~\cite{Murani_2004}, requiring a simple description by a response function. In contrast, no magnetic peak has been reported, other than the 0.6-{\AA}$^{-1}$ peak, hampering clarification of the spatial correlations. Different spatial-correlation models were also theoretically proposed, such as spin-orbit fluctuations with molecular V tetrahedra and one-dimensional (1D)-like chains~\cite{Hattori_2009, Fujimoto_2002}. 

% In this study and 
%
%\section{Experiments}
%

In this study, we performed INS experiments on powder samples of LiV$_2$O$_4$ using a state-of-the-art time-of-flight spectrometer with large-solid-angle detectors, which allows us to investigate the spin fluctuations in a wide $(Q,E)$ space.  
We used the direct geometry chopper spectrometer AMATERAS (BL14) at the MLF of the J-PARC spallation neutron source (Japan)~\cite{Nakajima_2011}. 
%To minimize the background, this spectrometer consists of a long upstream curved super-mirror guide and a vacuum scattering chamber with both sample and detectors. The detector coverage is as large as $5^{\circ}$ to $110^{\circ}$ in the horizontal plane and $\pm20^{\circ}$ in the vertical direction. 
The incident energy ($E_i$) was simultaneously set to 3.1, 7.7, 15, and 24 meV using the multi-$E_{i}$ technique, and the $E$ resolution under elastic conditions was approximately 2.0, 2.6, 3.6, and 4.5{\%} to $E_i$, respectively. The main disk chopper speed was fixed at 300 Hz. The data were obtained by the UTSUSEMI software provided by the MLF~\cite{Inamura_2010}. 
Scattering from the empty-container background measurements was subtracted, and the absolute intensities were obtained by normalization to measure the incoherent scattering intensity from the sample. 
%The accuracy is about $\pm10${\%}. 
%
A powder sample of LiV$_2$O$_4$ was synthesized by a solid-state reaction method~\cite{Kondo_1999}. Li is in natural abundance. Approximately 7.3 g of the sample was placed onto an aluminum foil and shaped into a hollow cylinder with a thickness of 3 mm and a diameter of 20 mm in order to mitigate the neutron-absorption effects of Li nuclei as much as possible. The cylinder was kept in the thin aluminum container with He exchange gas that was placed under a cold head in a He closed-cycle refrigerator. 

%
%\section{Results}
%
%detailed feature in a wider Q, E space

%
\begin{figure}[htbp]
\begin{center}
\includegraphics[width=0.95\linewidth, keepaspectratio]{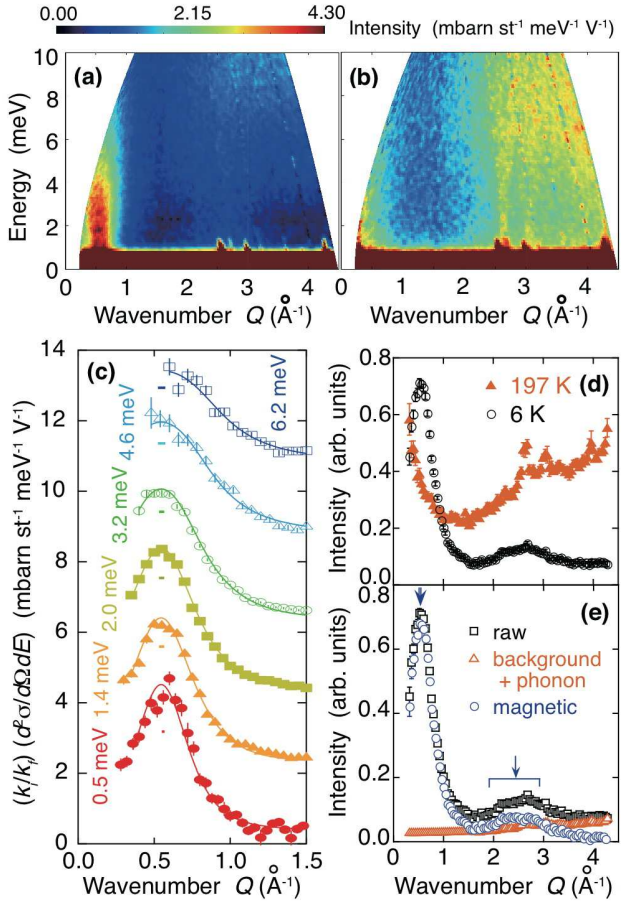}
\end{center}
\caption{\label{fig:data} (Color online)
(a), (b) Contour plots of the scattering intensity distribution in $(Q,E)$ space. 
(c) Constant-$E$ cross sections in the low-$Q$ range (symbols). From the bottom, $E_i$ = 3.1, 15, 15, 15, 24, and 24 meV. The averaged $E$ range was $\pm0.1$, $\pm0.3$, $\pm0.3$, $\pm0.3$, $\pm0.8$, and $\pm0.8$ meV. The vertical zero points are shifted by 2 mbarn$\cdot$st$^{-1}$$\cdot$meV$^{-1}$$\cdot$V$^{-1}$. The solid curves are fits of Eq.~(\ref{eq:chi''2}) to the data (see text). The tiny horizontal bars represent the $Q$ resolution, which can be neglected compared to the experimental line widths. 
(d) Constant-$E$ cross sections averaged in the $2.5\pm0.5$-meV range in (a) and (b). 
(e) Extraction of the magnetic component. The blue arrows indicate the magnetic peaks.}
\end{figure}

$Results.$---
%0.6-A^-1 peak
Figures~\ref{fig:data}(a) and \ref{fig:data}(b) show the observed scattering intensity distributions in $(Q,E)$ space. In the low-$Q$ range below 1.5 {\AA}$^{-1}$, which was previously reported~\cite{Lee_2001,Murani_2004}, magnetic scattering is observed with fountain-like $E$ evolution around 0.55 {\AA}$^{-1}$ at 6 K [Fig.~\ref{fig:data}(a)]. The scattering is paramagnetic around $0$ {\AA}$^{-1}$ at 197 K [Fig.~\ref{fig:data}(b)]. The constant-$E$ cross sections of the 6-K data are shown in Fig.~\ref{fig:data}(c). As $E$ increases, the scattering broadens in $Q$. 

%2.4-A^-1 peak
In addition, we searched for another unreported magnetic signal in the high-$Q$ range above 1.5 {\AA}$^{-1}$. Figure~\ref{fig:data}(d) shows the cross sections at 2.5 meV. This energy was selected to avoid the elastic tail and minimize phonon contamination. A broad and very weak signal appears between 2 and 3 {\AA}$^{-1}$ at 6 K, whereas strong phonon scattering with an intensity increasing with an increase in $Q$ is observed at 197 K. Thus, to examine whether the 6-K broad peak is magnetic in origin, we subtract the 197-K data and correct with a Bose population factor as the phonon component from the 6-K data, as shown in Fig.~\ref{fig:data}(e). In the low-$Q$ range, the phonon component was extrapolated from the high-$Q$ data by the $Q^2$ term with a constant background to avoid the tail of paramagnetic scattering at 197 K. After this subtraction, the 2.4-{\AA}$^{-1}$ broad peak still remained [open circles in Fig.~\ref{fig:data}(e)], indicating that another magnetic signal was found. 
%to avoid the tail of paramagnetic scattering in the low-$Q$ range below 1.5 {\AA}$^{-1}$ at 197 K

%
%\section{Analysis: Pines function}
%

%
\begin{figure}[htbp]
\begin{center}
\includegraphics[width=0.95\linewidth, keepaspectratio]{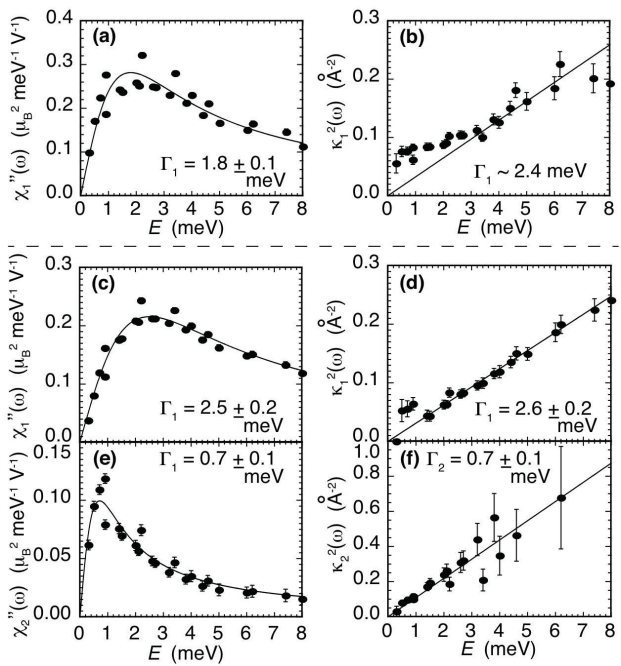}
\end{center}
\caption{\label{fig:fittings_Pines} Fitting results in the $n=1$ [(a)--(b)] and $n=2$ models [(c)--(f)]. Subfigures (a), (c), and (e) show the $E=\hbar\omega$ dependence of the dynamic susceptibilities at $Q_0$ and $\chi''_{j}(\omega)$, and panels (b), (d), and (f) show that of the $Q$-width parameters, $\kappa_{j}^{2}(\omega)$. The solid lines are fits of Eqs.~(\ref{eq:chi''2a}) and ~(\ref{eq:chi''2b}) to the data (see text). 
} 
\end{figure}

$Analyses.$---
We analyze the fountain-like $E$ evolution around 0.55 {\AA}$^{-1}$ at 6 K. The differential cross section of INS is proportional to the imaginary part $\chi''(Q,\omega)$ of the generalized magnetic susceptibility $\chi(Q,\omega)$, where $E=\hbar\omega$~\cite{Lovesey_text}. Following a standard way to describe the spin fluctuations in nearly AF metals such as Cr$_{0.95}$V$_{0.05}$ and La$_{2-x}$Sr$_{x}$CuO$_{4}$~\cite{Hayden_2000, Zha_1996}, we use  
\begin{equation}
\label{eq:chi}
\chi(Q,\omega) = \sum_{j=1}^{n}\chi_{j} \left[ 1+\frac{(Q-Q_{0})^{2}}{\kappa_{0}^{2}} - i \frac{\omega}{\Gamma_{j}} \right]^{-1}, 
\end{equation}
where $n$ is the number of spin fluctuation modes. This function corresponds to the expansion of the Lindhard function near the Fermi energy to describe a Fermi liquid~\cite{Hayden_2000, Zha_1996, Moriya_1970, Moriya_2010}. 
The imaginary part of Eq.~(\ref{eq:chi}) is described by the following useful form~\cite{Hayden_2000}: 
\begin{equation}
\label{eq:chi''2}
\chi''(Q,\omega) = \sum_{j=1}^{n} \chi''_{j}(\omega) \frac{\kappa_{0}^{4} + \kappa_{j}^{4}(\omega)}{\{\kappa_{0}^{2} + (Q-Q_{0})^{2}\}^{2} + \kappa_{j}^{4}(\omega)},  
\end{equation}
where the $\omega$ evolutions are separated into the susceptibility at $Q_0$, 
\begin{equation}
\label{eq:chi''2a}
\chi''_{j}(\omega) = \chi_{j} \frac{\omega\Gamma_{j}}{\omega^{2} + \Gamma_{j}^{2}}, 
\end{equation}
and the $Q$ width around $Q_0$, 
\begin{equation}
\label{eq:chi''2b}
\kappa_{j}^{2}(\omega) = \kappa_{0}^{2} \frac{\omega}{\Gamma_{j}}. 
\end{equation}
If Eq.~(\ref{eq:chi}) describes the data, an identical $\Gamma_{j}$ will be obtained for both the susceptibility and the $Q$-width parts. 

Figures~\ref{fig:fittings_Pines}(a) and \ref{fig:fittings_Pines}(b) show the fitting results for an $n=1$ model. The results coincide with Lee {\it et al.}'s INS report for both $\chi''_{1}(\omega)$ and $\Gamma_{1}$ [Fig.~\ref{fig:fittings_Pines}(a)]~\cite{Lee_2001}. 
However, $\kappa_{1}^{2}(\omega)$ is not proportional to $E$ in the low-$E$ region [Fig.~\ref{fig:fittings_Pines}(b)]. This is rather consistent with Murani {\it et al.}'s INS report and the $\mu$SR data, suggesting the coexistence of another slower component below 1 meV~\cite{Murani_2004, Kadono_2012}. 
Thus, we used an $n=2$ model, and the fitting results are shown in Figs.~\ref{fig:fittings_Pines}(c)--\ref{fig:fittings_Pines}(f). $\chi''_{1}(\omega)$, $\kappa_{1}^{2}(\omega)$, $\chi''_{2}(\omega)$, and $\kappa_{2}^{2}(\omega)$ are all fit well with $\Gamma_{1}=2.6$ meV and $\Gamma_{2}=0.7$ meV, where $\chi_{1}=0.43$ $\mu_{\rm B}^{2}$$\cdot$meV$^{-1}$$\cdot$V$^{-1}$, $\chi_{2}=0.20$ $\mu_{\rm B}^{2}$$\cdot$meV$^{-1}$$\cdot$V$^{-1}$, $Q_{0}=0.55$ {\AA}$^{-1}$, and $\kappa_{0}=0.28$ {\AA}$^{-1}$. 
%$\Gamma_{1}=2.6\pm0.2$ meV and $\Gamma_{2}=0.7\pm0.1$ meV, where $Q_{0}=0.55\pm0.005$ {\AA}$^{-1}$ and $\kappa_{0}=0.28\pm0.01$ {\AA}$^{-1}$

%
%\section{Analysis: spatial correlation}
%

%
\begin{figure*}[htbp]
\begin{center}
\includegraphics[width=0.90\linewidth, keepaspectratio]{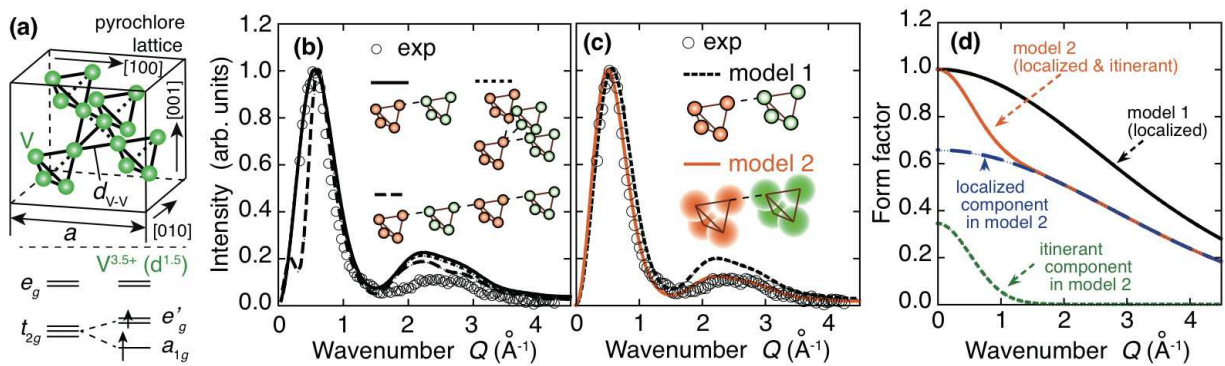}
\end{center}
\caption{\label{fig:fittings_mol}  
(Color online) Modeling of the dynamical spatial correlations. 
(a) Unit cell of the pyrochlore lattice (upper) and energy scheme of the V orbitals (lower). 
(b), (c) Comparisons of the experimental and model-calculated $Q$ dependencies of the intensities. All intensities are normalized at the maximum intensity $Q$. The experimental data (open circles) are identical to the magnetic scattering in Fig.~\ref{fig:data}(e). The curves were calculated in the inset models, where the orange and green spheres denote the dynamically fluctuating up and down spins, respectively. In (b) and (c), localized [model 1 in (d)] and modified [model 2 in (d)] magnetic form factors were used, respectively. 
(d) Magnetic form factors normalized at 0 {\AA}$^{-1}$. The broken and dotted curves resolve the model-2 form factor into localized and itinerant components, respectively. } 
\end{figure*}

Next, we analyze the spatial correlations of the fluctuations from the obtained $Q$ information. 
First, the value of $Q_{0}$ is equal to $0.72(2\pi/a) = 2\pi/\sqrt{2}a$, indicating the periodicity of $\sqrt{2}a=4d_{\rm V-V}$, where $a$ denotes the lattice constant of 8.24 {\AA} in the pyrochlore lattice~\cite{Matsushita_2005}, and $d_{\rm V-V}$ denotes the distance between the nearest-neighbor V sites, as shown in Fig.~\ref{fig:fittings_mol}(a). This strongly suggests that the fluctuations consist of AF bonds of four V atoms along the $\langle110\rangle$ direction. 
Second, the value of the $Q$-width $\kappa_{0}$ means $\sim5$ {\AA} of short correlation length. 
Third, as shown in Fig.~\ref{fig:data}(e), the $Q$ dependence of the intensity consists of a combination of a strong broad 0.6-{\AA}$^{-1}$ peak and a weak broad 2.4-{\AA}$^{-1}$ peak. This combination is identical to that of the 4-meV mode observed in another highly frustrated spinel-type insulator, GeCo$_2$O$_4$~\cite{Lashley_2008}, which is explained by AF di-tetrahedron spin correlations by single-crystal INS~\cite{Tomiyasu_2011_b}. Furthermore, this di-tetrahedron structure satisfies the first and second conditions. 
Thus, the fluctuations are most likely based on the di-tetrahedron in the spatial correlations. 

%In fact, 
We calculated the $Q$ dependence of the intensities for several di-tetrahedron-based models, and three of these models are shown in Fig.~\ref{fig:fittings_mol}(b), where a localized theoretical magnetic form factor was used~\cite{Iwata_1976}. As expected, every model roughly reproduces the 0.6-{\AA}$^{-1}$ and 2.4-{\AA}$^{-1}$ positions, and the di- and tetra-tetrahedron models are also consistent with the experimental data for the 0.6-{\AA}$^{-1}$ peak width. 

However, the calculated intensities of the 2.4-{\AA}$^{-1}$ peak are much stronger than the experimental data. To improve this, we incorporated the spatial expansion of the spin density distribution at each V site, considering that LiV$_2$O$_4$ exhibits metallic electrical resistivity. This itinerancy corresponds to the rapid decrease in the magnetic form factor in $Q$ space. 
Further, band calculations indicate that the Fermi level is mainly crossed by V 3$d$ $t_{2g}$ orbital bands, which roughly split into a localized $a_{1g}$ singlet and an itinerant $e_{g}'$ doublet via a small trigonal crystal field, as schematically shown in Fig.~\ref{fig:fittings_mol}(a)~\cite{Matsuno_1999,Animisov_1999}. 
Thus, for simplicity, we approximated the magnetic form factor by 
$\alpha f_{\rm localized}(Q) + (1-\alpha)\exp{(-Q^2/\Delta Q_{\rm itinerant})}$, 
where $f_{\rm localized}(Q)$ denotes the localized form factor~\cite{Iwata_1976} normalized at 0 {\AA}$^{-1}$, and $\alpha$ and $\Delta Q_{\rm itinerant}$ are determined to fit the experimental data. The spatial spin correlation was set to the di-tetrahedron. 

Figure~\ref{fig:fittings_mol}(c) shows a comparison among the experimental data, localized model 1, and model 2 with localization and itinerancy. The experimental data are well fit to model 2, which is much better than the model 1 with respect to the 2.4-{\AA}$^{-1}$ intensity and the 0.6-{\AA}$^{-1}$ peak profile. 
The model-2 fitting was obtained at $\alpha=0.66$ and $\Delta Q_{\rm itinerant}=0.74$ {\AA}$^{-1}$. The value of $\alpha$ indicates $\alpha:(1-\alpha)=1.0:0.5$, which is in excellent agreement with the Hund-rule filling of 1 $a_{1g}$ and 0.5 $e_{g}'$ electrons/V, as shown in Fig.~\ref{fig:fittings_mol}(a). The $\Delta Q_{\rm itinerant}$ value means 2.2 {\AA} of spatial distribution at half width at half maximum, which is much larger than $(d_{\rm V-V}/2)=1.5$ {\AA}. This indicates that the itinerant components of the nearest-neighbor V spins considerably overlap with each other in the di-tetrahedron. 

%
%\section{Discussion}
%
%HF, Fermi liquid, spin fluctuation, frustration
$Discussion.$---
%Now we discuss the aforementioned experiment and analysis results. 
The observed dynamical magnetic susceptibility is well described by a simple function for a Fermi liquid [Eq.~(\ref{eq:chi})] and is accompanied with the high itinerancy in the magnetic form factor. These facts verify the treatments of spin fluctuations in itinerant systems, such as a series of self-consistent renormalization theories, demonstrating that the spin-fluctuation channel dominates HF formation~\cite{Yushankhai_2007,Yushankhai_2008_a, Yushankhai_2010}. 
Further, the spin fluctuations are based on the AF di-tetrahedron in the spatial correlations. This strongly suggests that geometrical frustration causes the fluctuations as well as in GeCo$_2$O$_4$~\cite{Tomiyasu_2011_b}, and that the large HF entropy originates from the high degeneracy driven by frustration. 

%di-tetrahedron, frustration, spin-orbit fluctuations
Interestingly, the di-tetrahedron combines both theoretically proposed characteristics, the 1D-like chain along the $\langle110\rangle$ direction and the ferromagnetic tetrahedron units, which are accompanied by spin-orbit fluctuations~\cite{Fujimoto_2002, Hattori_2009}. 
In the former theory~\cite{Fujimoto_2002}, 1D correlations occur to release geometrical frustration, taking into account the fact that a pyrochlore lattice consists of 1D chains along the $\langle110\rangle$ directions, as shown in Fig.~\ref{fig:fittings_mol}(a). The V $t_{2g}$ orbitals hybridize with each other to form the 1D Hubbard chain with a periodicity of $4d_{\rm V-V}$. 
In the latter theory~\cite{Hattori_2009}, the ferromagnetic tetrahedra form to release frustration because a pyrochlore lattice is also regarded as the tetrahedra arranged in a face-centered-cubic lattice. The molecular orbital formation of the V$_4$ tetramer remarkably decreases the system energy. Further, the V$_4$ molecular orbital is half filled, which makes the inter-tetramer exchange interaction AF. 

%
%\section{Conclusions/Summary/Futures}
%
$Summary.$---
%In summary, 
We studied spin fluctuations over a wide $(Q,E)$ space in LiV$_2$O$_4$ by INS. The observed data can be described by a simple response function, a highly itinerant magnetic form factor, and AF di-tetrahedron-based spatial correlations. With these characteristics, the large HF entropy is attributable to frustration with spin-orbit fluctuations and remarkable orbital hybridization. Our study will promote future studies of novel quasiparticles as a prototype in the longstanding many-body problem.

\acknowledgments
We thank Dr. S. Iikubo for assisting with the sample preparation and Dr. M. Yokoyama for providing the preliminary neutron machine time. The neutron experiments were performed with the approval of J-PARC (2012A0146 and partially 2012P0202). This study was financially supported by Grants-in-Aid for Young Scientists (B) (22740209 and 26800174) and Priority Areas (22014001) from the MEXT of Japan. 
%, and Scientific Researches (A) (22244039) and (S) (21224008)

\bibliography{LiV2O4_3b_arXiv}

\begin{thebibliography}{31}
\expandafter\ifx\csname natexlab\endcsname\relax\def\natexlab#1{#1}\fi
\expandafter\ifx\csname bibnamefont\endcsname\relax
  \def\bibnamefont#1{#1}\fi
\expandafter\ifx\csname bibfnamefont\endcsname\relax
  \def\bibfnamefont#1{#1}\fi
\expandafter\ifx\csname citenamefont\endcsname\relax
  \def\citenamefont#1{#1}\fi
\expandafter\ifx\csname url\endcsname\relax
  \def\url#1{\texttt{#1}}\fi
\expandafter\ifx\csname urlprefix\endcsname\relax\def\urlprefix{URL }\fi
\providecommand{\bibinfo}[2]{#2}
\providecommand{\eprint}[2][]{\url{#2}}

\bibitem[{\citenamefont{Landau and Lifshitz}(Pergamon, Oxford,
  1981)}]{Landau_text}
\bibinfo{author}{\bibfnamefont{L.~D.} \bibnamefont{Landau}} \bibnamefont{and}
  \bibinfo{author}{\bibfnamefont{E.~M.} \bibnamefont{Lifshitz}},
  \bibinfo{journal}{{\it Course of Theoretical Physics Vol. 9, Statistical
  Physics Part 2}}  (\bibinfo{year}{Pergamon, Oxford, 1981}).

\bibitem[{\citenamefont{Kondo et~al.}(1997)\citenamefont{Kondo, Johnston,
  Swenson, Borsa, Mahajan, Miller, Gu, Goldman, Maple, Gajewski
  et~al.}}]{Kondo_1997}
\bibinfo{author}{\bibfnamefont{S.}~\bibnamefont{Kondo}},
  \bibinfo{author}{\bibfnamefont{D.~C.} \bibnamefont{Johnston}},
  \bibinfo{author}{\bibfnamefont{C.~A.} \bibnamefont{Swenson}},
  \bibinfo{author}{\bibfnamefont{F.}~\bibnamefont{Borsa}},
  \bibinfo{author}{\bibfnamefont{A.~V.} \bibnamefont{Mahajan}},
  \bibinfo{author}{\bibfnamefont{L.~L.} \bibnamefont{Miller}},
  \bibinfo{author}{\bibfnamefont{T.}~\bibnamefont{Gu}},
  \bibinfo{author}{\bibfnamefont{A.~I.} \bibnamefont{Goldman}},
  \bibinfo{author}{\bibfnamefont{M.~B.} \bibnamefont{Maple}},
  \bibinfo{author}{\bibfnamefont{D.~A.} \bibnamefont{Gajewski}},
  \bibnamefont{et~al.}, \bibinfo{journal}{Phys. Rev. Lett.}
  \textbf{\bibinfo{volume}{78}}, \bibinfo{pages}{3729} (\bibinfo{year}{1997}).

\bibitem[{\citenamefont{Urano et~al.}(2000)\citenamefont{Urano, Nohara, Kondo,
  Sakai, Takagi, Shiraki, and Okubo}}]{Urano_2000}
\bibinfo{author}{\bibfnamefont{C.}~\bibnamefont{Urano}},
  \bibinfo{author}{\bibfnamefont{M.}~\bibnamefont{Nohara}},
  \bibinfo{author}{\bibfnamefont{S.}~\bibnamefont{Kondo}},
  \bibinfo{author}{\bibfnamefont{F.}~\bibnamefont{Sakai}},
  \bibinfo{author}{\bibfnamefont{H.}~\bibnamefont{Takagi}},
  \bibinfo{author}{\bibfnamefont{T.}~\bibnamefont{Shiraki}}, \bibnamefont{and}
  \bibinfo{author}{\bibfnamefont{T.}~\bibnamefont{Okubo}},
  \bibinfo{journal}{Phys. Rev. Lett.} \textbf{\bibinfo{volume}{85}},
  \bibinfo{pages}{1052} (\bibinfo{year}{2000}).

\bibitem[{\citenamefont{Matsushita et~al.}(2005)\citenamefont{Matsushita, Ueda,
  and Ueda}}]{Matsushita_2005}
\bibinfo{author}{\bibfnamefont{Y.}~\bibnamefont{Matsushita}},
  \bibinfo{author}{\bibfnamefont{H.}~\bibnamefont{Ueda}}, \bibnamefont{and}
  \bibinfo{author}{\bibfnamefont{Y.}~\bibnamefont{Ueda}},
  \bibinfo{journal}{Nature} \textbf{\bibinfo{volume}{4}}, \bibinfo{pages}{845}
  (\bibinfo{year}{2005}).

\bibitem[{\citenamefont{Das et~al.}(2007)\citenamefont{Das, Zong, Niazi,
  Ellern, Yan, and Johnston}}]{Das_2007}
\bibinfo{author}{\bibfnamefont{S.}~\bibnamefont{Das}},
  \bibinfo{author}{\bibfnamefont{X.}~\bibnamefont{Zong}},
  \bibinfo{author}{\bibfnamefont{A.}~\bibnamefont{Niazi}},
  \bibinfo{author}{\bibfnamefont{A.}~\bibnamefont{Ellern}},
  \bibinfo{author}{\bibfnamefont{J.~Q.} \bibnamefont{Yan}}, \bibnamefont{and}
  \bibinfo{author}{\bibfnamefont{D.~C.} \bibnamefont{Johnston}},
  \bibinfo{journal}{Phys. Rev. B} \textbf{\bibinfo{volume}{76}},
  \bibinfo{pages}{054418} (\bibinfo{year}{2007}).

\bibitem[{\citenamefont{Wada et~al.}(1987)\citenamefont{Wada, Nakamura, Fukami,
  Yoshimura, Shiga, and Nakamura}}]{Wada_1987}
\bibinfo{author}{\bibfnamefont{H.}~\bibnamefont{Wada}},
  \bibinfo{author}{\bibfnamefont{H.}~\bibnamefont{Nakamura}},
  \bibinfo{author}{\bibfnamefont{E.}~\bibnamefont{Fukami}},
  \bibinfo{author}{\bibfnamefont{K.}~\bibnamefont{Yoshimura}},
  \bibinfo{author}{\bibfnamefont{M.}~\bibnamefont{Shiga}}, \bibnamefont{and}
  \bibinfo{author}{\bibfnamefont{Y.}~\bibnamefont{Nakamura}},
  \bibinfo{journal}{J. Mag. Mag. Mat.} \textbf{\bibinfo{volume}{70}},
  \bibinfo{pages}{17} (\bibinfo{year}{1987}).

\bibitem[{\citenamefont{Lee et~al.}(2001)\citenamefont{Lee, Qiu, Broholm, Ueda,
  and Rush}}]{Lee_2001}
\bibinfo{author}{\bibfnamefont{S.-H.} \bibnamefont{Lee}},
  \bibinfo{author}{\bibfnamefont{Y.}~\bibnamefont{Qiu}},
  \bibinfo{author}{\bibfnamefont{C.}~\bibnamefont{Broholm}},
  \bibinfo{author}{\bibfnamefont{Y.}~\bibnamefont{Ueda}}, \bibnamefont{and}
  \bibinfo{author}{\bibfnamefont{J.~J.} \bibnamefont{Rush}},
  \bibinfo{journal}{Phys. Rev. Lett.} \textbf{\bibinfo{volume}{86}},
  \bibinfo{pages}{5554} (\bibinfo{year}{2001}).

\bibitem[{\citenamefont{Shimizu et~al.}(2012)\citenamefont{Shimizu, Takeda,
  Tanaka, Itoh, Niitaka, and Takagi}}]{Shimizu_2012}
\bibinfo{author}{\bibfnamefont{Y.}~\bibnamefont{Shimizu}},
  \bibinfo{author}{\bibfnamefont{H.}~\bibnamefont{Takeda}},
  \bibinfo{author}{\bibfnamefont{M.}~\bibnamefont{Tanaka}},
  \bibinfo{author}{\bibfnamefont{M.}~\bibnamefont{Itoh}},
  \bibinfo{author}{\bibfnamefont{S.}~\bibnamefont{Niitaka}}, \bibnamefont{and}
  \bibinfo{author}{\bibfnamefont{H.}~\bibnamefont{Takagi}},
  \bibinfo{journal}{Nature Comm.} \textbf{\bibinfo{volume}{3}},
  \bibinfo{pages}{981} (\bibinfo{year}{2012}).

\bibitem[{\citenamefont{Kadono et~al.}(2012)\citenamefont{Kadono, Koda,
  Higemoto, Ohishi, Ueda, Urano, Kondo, Nohara, and Takagi}}]{Kadono_2012}
\bibinfo{author}{\bibfnamefont{R.}~\bibnamefont{Kadono}},
  \bibinfo{author}{\bibfnamefont{A.}~\bibnamefont{Koda}},
  \bibinfo{author}{\bibfnamefont{W.}~\bibnamefont{Higemoto}},
  \bibinfo{author}{\bibfnamefont{K.}~\bibnamefont{Ohishi}},
  \bibinfo{author}{\bibfnamefont{H.}~\bibnamefont{Ueda}},
  \bibinfo{author}{\bibfnamefont{C.}~\bibnamefont{Urano}},
  \bibinfo{author}{\bibfnamefont{S.}~\bibnamefont{Kondo}},
  \bibinfo{author}{\bibfnamefont{M.}~\bibnamefont{Nohara}}, \bibnamefont{and}
  \bibinfo{author}{\bibfnamefont{H.}~\bibnamefont{Takagi}},
  \bibinfo{journal}{J. Phys. Soc. Jpn.} \textbf{\bibinfo{volume}{81}},
  \bibinfo{pages}{014709} (\bibinfo{year}{2012}).

\bibitem[{\citenamefont{Yushankhai et~al.}(2007)\citenamefont{Yushankhai,
  Yaresko, Fulde, and Thalmeier}}]{Yushankhai_2007}
\bibinfo{author}{\bibfnamefont{V.}~\bibnamefont{Yushankhai}},
  \bibinfo{author}{\bibfnamefont{A.}~\bibnamefont{Yaresko}},
  \bibinfo{author}{\bibfnamefont{P.}~\bibnamefont{Fulde}}, \bibnamefont{and}
  \bibinfo{author}{\bibfnamefont{P.}~\bibnamefont{Thalmeier}},
  \bibinfo{journal}{Phys. Rev. B} \textbf{\bibinfo{volume}{76}},
  \bibinfo{pages}{085111} (\bibinfo{year}{2007}).

\bibitem[{\citenamefont{J$\ddot{\rm o}$nsson
  et~al.}(2007)\citenamefont{J$\ddot{\rm o}$nsson, Takenaka, Niitaka, Sasagawa,
  Sugai, and Takagi}}]{Jonsson_2007}
\bibinfo{author}{\bibfnamefont{P.~E.} \bibnamefont{J$\ddot{\rm o}$nsson}},
  \bibinfo{author}{\bibfnamefont{K.}~\bibnamefont{Takenaka}},
  \bibinfo{author}{\bibfnamefont{S.}~\bibnamefont{Niitaka}},
  \bibinfo{author}{\bibfnamefont{T.}~\bibnamefont{Sasagawa}},
  \bibinfo{author}{\bibfnamefont{S.}~\bibnamefont{Sugai}}, \bibnamefont{and}
  \bibinfo{author}{\bibfnamefont{H.}~\bibnamefont{Takagi}},
  \bibinfo{journal}{Phys. Rev. Lett.} \textbf{\bibinfo{volume}{99}},
  \bibinfo{pages}{167402} (\bibinfo{year}{2007}).

\bibitem[{\citenamefont{Shimoyamada et~al.}(2006)\citenamefont{Shimoyamada,
  Tsuda, Ishizaka, Kiss, Shimojima, Togashi, Watanabe, Zhang, Chen, Matsushita
  et~al.}}]{Shimoyamada_2006}
\bibinfo{author}{\bibfnamefont{A.}~\bibnamefont{Shimoyamada}},
  \bibinfo{author}{\bibfnamefont{S.}~\bibnamefont{Tsuda}},
  \bibinfo{author}{\bibfnamefont{K.}~\bibnamefont{Ishizaka}},
  \bibinfo{author}{\bibfnamefont{T.}~\bibnamefont{Kiss}},
  \bibinfo{author}{\bibfnamefont{T.}~\bibnamefont{Shimojima}},
  \bibinfo{author}{\bibfnamefont{T.}~\bibnamefont{Togashi}},
  \bibinfo{author}{\bibfnamefont{S.}~\bibnamefont{Watanabe}},
  \bibinfo{author}{\bibfnamefont{C.~Q.} \bibnamefont{Zhang}},
  \bibinfo{author}{\bibfnamefont{C.~T.} \bibnamefont{Chen}},
  \bibinfo{author}{\bibfnamefont{Y.}~\bibnamefont{Matsushita}},
  \bibnamefont{et~al.}, \bibinfo{journal}{Phys. Rev. Lett.}
  \textbf{\bibinfo{volume}{96}}, \bibinfo{pages}{026403}
  (\bibinfo{year}{2006}).

\bibitem[{\citenamefont{Arita et~al.}(2007)\citenamefont{Arita, Held,
  Lukoyanov, and Anisimov}}]{Arita_2007}
\bibinfo{author}{\bibfnamefont{R.}~\bibnamefont{Arita}},
  \bibinfo{author}{\bibfnamefont{K.}~\bibnamefont{Held}},
  \bibinfo{author}{\bibfnamefont{A.~V.} \bibnamefont{Lukoyanov}},
  \bibnamefont{and} \bibinfo{author}{\bibfnamefont{V.~I.}
  \bibnamefont{Anisimov}}, \bibinfo{journal}{Phys. Rev. Lett}
  \textbf{\bibinfo{volume}{98}}, \bibinfo{pages}{166402}
  (\bibinfo{year}{2007}).

\bibitem[{\citenamefont{Murani et~al.}(2004)\citenamefont{Murani, Krimmel,
  Stewart, Smith, Strobel, Loidl, and Ibarra-Palos}}]{Murani_2004}
\bibinfo{author}{\bibfnamefont{A.~P.} \bibnamefont{Murani}},
  \bibinfo{author}{\bibfnamefont{A.}~\bibnamefont{Krimmel}},
  \bibinfo{author}{\bibfnamefont{J.~R.} \bibnamefont{Stewart}},
  \bibinfo{author}{\bibfnamefont{M.}~\bibnamefont{Smith}},
  \bibinfo{author}{\bibfnamefont{P.}~\bibnamefont{Strobel}},
  \bibinfo{author}{\bibfnamefont{A.}~\bibnamefont{Loidl}}, \bibnamefont{and}
  \bibinfo{author}{\bibfnamefont{A.}~\bibnamefont{Ibarra-Palos}},
  \bibinfo{journal}{J. Phys.: Condens. Matter} \textbf{\bibinfo{volume}{16}},
  \bibinfo{pages}{S607} (\bibinfo{year}{2004}).

\bibitem[{\citenamefont{Hattori and Tsunetsugu}(2009)}]{Hattori_2009}
\bibinfo{author}{\bibfnamefont{K.}~\bibnamefont{Hattori}} \bibnamefont{and}
  \bibinfo{author}{\bibfnamefont{H.}~\bibnamefont{Tsunetsugu}},
  \bibinfo{journal}{Phys. Rev. B} \textbf{\bibinfo{volume}{79}},
  \bibinfo{pages}{035115} (\bibinfo{year}{2009}).

\bibitem[{\citenamefont{Fujimoto}(2002)}]{Fujimoto_2002}
\bibinfo{author}{\bibfnamefont{S.}~\bibnamefont{Fujimoto}},
  \bibinfo{journal}{Phys. Rev. B} \textbf{\bibinfo{volume}{65}},
  \bibinfo{pages}{155108} (\bibinfo{year}{2002}).

\bibitem[{\citenamefont{Nakajima et~al.}(2011)\citenamefont{Nakajima,
  Ohira-Kawamura, Kikuchi, Nakamura, Kajimoto, Inamura, Takahashi, Aizawa,
  Suzuya, Shibata et~al.}}]{Nakajima_2011}
\bibinfo{author}{\bibfnamefont{K.}~\bibnamefont{Nakajima}},
  \bibinfo{author}{\bibfnamefont{S.}~\bibnamefont{Ohira-Kawamura}},
  \bibinfo{author}{\bibfnamefont{T.}~\bibnamefont{Kikuchi}},
  \bibinfo{author}{\bibfnamefont{M.}~\bibnamefont{Nakamura}},
  \bibinfo{author}{\bibfnamefont{R.}~\bibnamefont{Kajimoto}},
  \bibinfo{author}{\bibfnamefont{Y.}~\bibnamefont{Inamura}},
  \bibinfo{author}{\bibfnamefont{N.}~\bibnamefont{Takahashi}},
  \bibinfo{author}{\bibfnamefont{K.}~\bibnamefont{Aizawa}},
  \bibinfo{author}{\bibfnamefont{K.}~\bibnamefont{Suzuya}},
  \bibinfo{author}{\bibfnamefont{K.}~\bibnamefont{Shibata}},
  \bibnamefont{et~al.}, \bibinfo{journal}{J. Phys. Soc. Jpn.}
  \textbf{\bibinfo{volume}{80}}, \bibinfo{pages}{SB028} (\bibinfo{year}{2011}).

\bibitem[{\citenamefont{Inamura et~al.}(2010)\citenamefont{Inamura, Nakajima,
  Kajimoto, Nakatani, Arai, Otomo, Suzuki, So, and Park}}]{Inamura_2010}
\bibinfo{author}{\bibfnamefont{Y.}~\bibnamefont{Inamura}},
  \bibinfo{author}{\bibfnamefont{K.}~\bibnamefont{Nakajima}},
  \bibinfo{author}{\bibfnamefont{R.}~\bibnamefont{Kajimoto}},
  \bibinfo{author}{\bibfnamefont{T.}~\bibnamefont{Nakatani}},
  \bibinfo{author}{\bibfnamefont{M.}~\bibnamefont{Arai}},
  \bibinfo{author}{\bibfnamefont{T.}~\bibnamefont{Otomo}},
  \bibinfo{author}{\bibfnamefont{J.}~\bibnamefont{Suzuki}},
  \bibinfo{author}{\bibfnamefont{J.~Y.} \bibnamefont{So}}, \bibnamefont{and}
  \bibinfo{author}{\bibfnamefont{J.~G.} \bibnamefont{Park}},
  \bibinfo{journal}{Proc. 19th Meet. Int. Collaboration of Advanced Neutron
  Sources (PSI-Proceedings 10-01)}  (\bibinfo{year}{2010}).

\bibitem[{\citenamefont{Kondo et~al.}(1999)\citenamefont{Kondo, Johnston, and
  Miller}}]{Kondo_1999}
\bibinfo{author}{\bibfnamefont{S.}~\bibnamefont{Kondo}},
  \bibinfo{author}{\bibfnamefont{D.~C.} \bibnamefont{Johnston}},
  \bibnamefont{and} \bibinfo{author}{\bibfnamefont{L.~L.}
  \bibnamefont{Miller}}, \bibinfo{journal}{Phys. Rev. B}
  \textbf{\bibinfo{volume}{59}}, \bibinfo{pages}{2609} (\bibinfo{year}{1999}).

\bibitem[{\citenamefont{Lovesey}(Oxford University Press, 1984)}]{Lovesey_text}
\bibinfo{author}{\bibfnamefont{S.~W.} \bibnamefont{Lovesey}},
  \bibinfo{journal}{{\it Theory of Neutron Scattering from Condensed Matter}}
  (\bibinfo{year}{Oxford University Press, 1984}).

\bibitem[{\citenamefont{Hayden et~al.}(2000)\citenamefont{Hayden, Doubble,
  Aeppli, Perring, and Fawcett}}]{Hayden_2000}
\bibinfo{author}{\bibfnamefont{S.~M.} \bibnamefont{Hayden}},
  \bibinfo{author}{\bibfnamefont{R.}~\bibnamefont{Doubble}},
  \bibinfo{author}{\bibfnamefont{G.}~\bibnamefont{Aeppli}},
  \bibinfo{author}{\bibfnamefont{T.~G.} \bibnamefont{Perring}},
  \bibnamefont{and} \bibinfo{author}{\bibfnamefont{E.}~\bibnamefont{Fawcett}},
  \bibinfo{journal}{Phys. Rev. Lett.} \textbf{\bibinfo{volume}{84}},
  \bibinfo{pages}{999} (\bibinfo{year}{2000}).

\bibitem[{\citenamefont{Zha et~al.}(1996)\citenamefont{Zha, Barzykin, and
  Pines}}]{Zha_1996}
\bibinfo{author}{\bibfnamefont{Y.}~\bibnamefont{Zha}},
  \bibinfo{author}{\bibfnamefont{V.}~\bibnamefont{Barzykin}}, \bibnamefont{and}
  \bibinfo{author}{\bibfnamefont{D.}~\bibnamefont{Pines}},
  \bibinfo{journal}{Phys. Rev. B} \textbf{\bibinfo{volume}{54}},
  \bibinfo{pages}{7561} (\bibinfo{year}{1996}).

\bibitem[{\citenamefont{Moriya}(1970)}]{Moriya_1970}
\bibinfo{author}{\bibfnamefont{T.}~\bibnamefont{Moriya}},
  \bibinfo{journal}{Phys. Rev. Lett.} \textbf{\bibinfo{volume}{24}},
  \bibinfo{pages}{1433} (\bibinfo{year}{1970}).

\bibitem[{\citenamefont{Moriya and Ueda}(2000)}]{Moriya_2010}
\bibinfo{author}{\bibfnamefont{T.}~\bibnamefont{Moriya}} \bibnamefont{and}
  \bibinfo{author}{\bibfnamefont{K.}~\bibnamefont{Ueda}},
  \bibinfo{journal}{Adv. Phys.} \textbf{\bibinfo{volume}{49}},
  \bibinfo{pages}{555} (\bibinfo{year}{2000}).

\bibitem[{\citenamefont{Lashley et~al.}(2008)\citenamefont{Lashley, Stevens,
  Crawford, Boerio-Goates, Woodfield, Qiu, Lynn, Goddard, and
  Fisher}}]{Lashley_2008}
\bibinfo{author}{\bibfnamefont{J.~C.} \bibnamefont{Lashley}},
  \bibinfo{author}{\bibfnamefont{R.}~\bibnamefont{Stevens}},
  \bibinfo{author}{\bibfnamefont{M.~K.} \bibnamefont{Crawford}},
  \bibinfo{author}{\bibfnamefont{J.}~\bibnamefont{Boerio-Goates}},
  \bibinfo{author}{\bibfnamefont{B.~F.} \bibnamefont{Woodfield}},
  \bibinfo{author}{\bibfnamefont{Y.}~\bibnamefont{Qiu}},
  \bibinfo{author}{\bibfnamefont{J.~W.} \bibnamefont{Lynn}},
  \bibinfo{author}{\bibfnamefont{P.~A.} \bibnamefont{Goddard}},
  \bibnamefont{and} \bibinfo{author}{\bibfnamefont{R.~A.}
  \bibnamefont{Fisher}}, \bibinfo{journal}{Phys. Rev. B}
  \textbf{\bibinfo{volume}{78}}, \bibinfo{pages}{104406}
  (\bibinfo{year}{2008}).

\bibitem[{\citenamefont{Tomiyasu et~al.}(2011)\citenamefont{Tomiyasu, Crawford,
  Adroja, Manuel, Tominaga, Hara, Sato, Watanabe, Ikeda, Lynn
  et~al.}}]{Tomiyasu_2011_b}
\bibinfo{author}{\bibfnamefont{K.}~\bibnamefont{Tomiyasu}},
  \bibinfo{author}{\bibfnamefont{M.~K.} \bibnamefont{Crawford}},
  \bibinfo{author}{\bibfnamefont{D.~T.} \bibnamefont{Adroja}},
  \bibinfo{author}{\bibfnamefont{P.}~\bibnamefont{Manuel}},
  \bibinfo{author}{\bibfnamefont{A.}~\bibnamefont{Tominaga}},
  \bibinfo{author}{\bibfnamefont{S.}~\bibnamefont{Hara}},
  \bibinfo{author}{\bibfnamefont{H.}~\bibnamefont{Sato}},
  \bibinfo{author}{\bibfnamefont{T.}~\bibnamefont{Watanabe}},
  \bibinfo{author}{\bibfnamefont{S.~I.} \bibnamefont{Ikeda}},
  \bibinfo{author}{\bibfnamefont{J.~W.} \bibnamefont{Lynn}},
  \bibnamefont{et~al.}, \bibinfo{journal}{Phys. Rev. B}
  \textbf{\bibinfo{volume}{84}}, \bibinfo{pages}{054405}
  (\bibinfo{year}{2011}).

\bibitem[{\citenamefont{Iwata}(1977)}]{Iwata_1976}
\bibinfo{author}{\bibfnamefont{M.}~\bibnamefont{Iwata}},
  \bibinfo{journal}{Acta. Cryst.} \textbf{\bibinfo{volume}{B33}},
  \bibinfo{pages}{59} (\bibinfo{year}{1977}).

\bibitem[{\citenamefont{Matsuno et~al.}(1999)\citenamefont{Matsuno, Fujimori,
  and Mattheiss}}]{Matsuno_1999}
\bibinfo{author}{\bibfnamefont{J.}~\bibnamefont{Matsuno}},
  \bibinfo{author}{\bibfnamefont{A.}~\bibnamefont{Fujimori}}, \bibnamefont{and}
  \bibinfo{author}{\bibfnamefont{L.~F.} \bibnamefont{Mattheiss}},
  \bibinfo{journal}{Phys. Rev. B} \textbf{\bibinfo{volume}{60}},
  \bibinfo{pages}{1607} (\bibinfo{year}{1999}).

\bibitem[{\citenamefont{Anisimov et~al.}(1999)\citenamefont{Anisimov, Korotin,
  Z$\ddot{\rm o}$lfl, Pruschke, Hur, and Rice}}]{Animisov_1999}
\bibinfo{author}{\bibfnamefont{V.~I.} \bibnamefont{Anisimov}},
  \bibinfo{author}{\bibfnamefont{M.~A.} \bibnamefont{Korotin}},
  \bibinfo{author}{\bibfnamefont{M.}~\bibnamefont{Z$\ddot{\rm o}$lfl}},
  \bibinfo{author}{\bibfnamefont{T.}~\bibnamefont{Pruschke}},
  \bibinfo{author}{\bibfnamefont{K.~L.} \bibnamefont{Hur}}, \bibnamefont{and}
  \bibinfo{author}{\bibfnamefont{T.~M.} \bibnamefont{Rice}},
  \bibinfo{journal}{Phys. Rev. Lett.} \textbf{\bibinfo{volume}{83}},
  \bibinfo{pages}{364} (\bibinfo{year}{1999}).

\bibitem[{\citenamefont{Yushankhai et~al.}(2008)\citenamefont{Yushankhai,
  Thalmeier, and Takimoto}}]{Yushankhai_2008_a}
\bibinfo{author}{\bibfnamefont{V.}~\bibnamefont{Yushankhai}},
  \bibinfo{author}{\bibfnamefont{P.}~\bibnamefont{Thalmeier}},
  \bibnamefont{and} \bibinfo{author}{\bibfnamefont{T.}~\bibnamefont{Takimoto}},
  \bibinfo{journal}{Phys. Rev. B} \textbf{\bibinfo{volume}{77}},
  \bibinfo{pages}{125126} (\bibinfo{year}{2008}).

\bibitem[{\citenamefont{Yushankhai et~al.}(2010)\citenamefont{Yushankhai,
  Takimoto, and Thalmeier}}]{Yushankhai_2010}
\bibinfo{author}{\bibfnamefont{V.}~\bibnamefont{Yushankhai}},
  \bibinfo{author}{\bibfnamefont{T.}~\bibnamefont{Takimoto}}, \bibnamefont{and}
  \bibinfo{author}{\bibfnamefont{P.}~\bibnamefont{Thalmeier}},
  \bibinfo{journal}{Phys. Rev. B} \textbf{\bibinfo{volume}{82}},
  \bibinfo{pages}{085112} (\bibinfo{year}{2010}).

\end{thebibliography}

\end{document}